%% file: borrett-hmg-revised.tex
\documentclass[preprint, 5p,authoryear]{elsarticle}
\journal{Ecological Modelling}
\usepackage{amsmath,amssymb,amsthm,amsfonts}
\usepackage[]{graphicx}
\usepackage{lineno,setspace}
\usepackage[pdftitle={Network Homogenization}, pdfauthor={Stuart
  R. Borrett}, pdfkeywords={ecology, network, indirect effects,
  environ, NEA, holoecology, network homogenization},
pdfpagemode=UseOutlines, bookmarks=T, colorlinks, linkcolor=blue,
citecolor=blue, pdfstartview=FitH, urlcolor=red]{hyperref}

\def\citeapos#1{\citeauthor{#1}'s (\citeyear{#1})}

\begin{document} 
\begin{frontmatter}
\title{Evidence for Resource Homogenization
    in 50 Trophic Ecosystem Networks}

\author{S.R.~Borrett\corref{cor1}}
\ead{borretts@uncw.edu}

\author{A.K.~Salas}
\ead{aks2515@uncw.edu}

\cortext[cor1]{Corresponding author. Tel. 910.962.2411; fax: 910.962.4066}

\address{Department of Biology \& Marine Biology and Center for Marine
  Science, University
  of North Carolina Wilmington, 601 S.\ College Rd., Wilmington, 28403
  NC, USA}

\begin{abstract}
  Connectivity patterns of ecological elements are often the core
  concern of ecologists working at multiple levels of organization
  (e.g., populations, communities, ecosystems, and landscapes) because
  these patterns often reflect the forces shaping the system's
  development as well as constraining their operation.
  One reason these patterns of direct connections are critical is that
  they establish the pathways through which elements influence
  each other indirectly.  Here, we tested a hypothesized consequence of
  connectivity in ecosystems: the homogenization of resource
  distributions in flow networks.  Specifically, we tested the
  generality of the systems ecology hypothesis of resource
  homogenization in 50 empirically derived trophic ecosystem models
  representing 35 distinct ecosystems.  We applied Network Environ
  Analysis (NEA) to calculate resource homogenization for these
  models, where homogenization is defined as the ratio of the
  coefficient of variation of the direct flow intensity matrix
  ($CV(\mathbf{G})$) to the covariance of the integral flow intensity
  matrix ($CV(\mathbf{N})$).  A ratio greater than unity indicates the
  presence of homogenization.  We also tested the hypotheses that
  homogenization increases with system size, connectance, and cycling.
  We further evaluated the robustness of our results in two ways.
  First, we verified the close correspondence between the input- and
  output-oriented homogenization values to ensure that our results
  were not biased by our decision to focus on the output orientation.
  Second, we conducted a Monte Carlo based uncertainty analysis to
  determine the robustness of our results to $\pm$5\% error introduced
  into the original flow matrices for each model.  Our results show
  that resource homogenization occurs universally in the 50 ecosystem
  models tested, with values ranging from 1.04 to 1.97 and a median of
  1.61.  However, our results do not support the hypothesized
  relationship between network homogenization and system size and
  connectance, as the results of the linear regressions are
  insignificant.  Further, there is only weak support for the positive
  relationship between homogenization and cycling.  We confirm that
  our results are not biased by using the output-oriented
  homogenization values instead of the input-oriented values because
  there is a significant linear regression between the two types of
  homogenization ($r^2 = 0.38$, $p < 0.001$) and the values are well
  correlated ($S = 8,054$, $\rho = 0.61 $, $p < 0.001$).  Finally, we
  found that our results are robust to $\pm$5\% error in the flow
  matrices.  The error in the homogenization values was less than the
  error introduced into the models and ranged from a minimum of
  0.24\% to a maximum of 1.5\% with a median value of 0.58\%.  The
  error did not change the qualitative interpretation of the
  homogenization values.  In conclusion, we found strong support for
  the resource homogenization hypothesis in 50 empirically derived
  ecosystem models.
\end{abstract}


\end{frontmatter}
\begin{spacing}{1}

\newpage
\begin{quote}
  ``It is a recognized principle of ecology that the interactions of
  organisms and their environment are reciprocal'' \cite{redfield58}
\end{quote}

\section{Introduction}
Connectivity is a core concept in ecology.  This is evident in
\citeapos{redfield58} opening quote, but other examples include
\citeapos{darwin59} entangled bank metaphor, exploiter--victim spatial
connectivity \citep{holland08}, landscape ecology \citep{urban01}, and
community interactions \citep{jordan03}. John Muir solidified this
idea in the public imagination with his statement that ``When we try
to pick out anything by itself, we find that it is bound fast by a
thousand invisible cords that cannot be broken to everything in the
universe''\citep[as quoted by][p. 291]{fox81}.

Too often this principle idea is dismissively summarized as:
``Everything is connected to everything else''. As \citet{peters91}
pointed out, this statement is vacuous; there is no way to determine
if this statement is true or false.  Perhaps more importantly,
however, it misses the fundamental point.  In ecological systems, as
well as other types of complex systems, \emph{how} organisms and their
environmental components are connected is what is interesting.  For
example, \citet{jordano03_invariant} suggest that the nestedness
pattern commonly found in mutualistic networks may facilitate
coevolutionary interactions.  \citet{dunne02} found increasing
food web connectivity tends to increase ecosystem robustness to
biodiversity loss, but this effect was modulated in part by the pattern
of node degree distributions.  These connectivity patterns reflect the
processes that create and constrain a system's development, its function
and the services it provides.  In this paper, we investigated a
hypothesized consequence of network connectivity in ecosystems, the
systems ecology hypothesis that ``the action of networks makes the
distribution of resources more uniform'', which is termed
\emph{network homogenization} \citep{patten90_cycles, fath99_homo,
  jorgensen07_newecology_conect}.


The network homogenization hypothesis emerged from the development and
application of Network Environ Analysis (NEA), one type of ecological
network analysis derived from the application and extension of
\citeapos{leontief66} economic input--output analysis \citep[see
  review by ][]{fath99_review}.  \citet{patten90_cycles} noticed that
when they applied NEA to ecosystem models, the distribution of
integral flow intensities (the combination of boundary, direct, and
indirect flow intensities) tended to be more even than the
distribution of direct flow intensities alone.  This change in
distribution was hypothesized to be a consequence of the indirect
relationships that depend in part on the connectivity pattern
\citep{borrett07_jtb,borrett10_idd}.  

\citet{fath99_homo} introduced a method of
quantifying network homogenization that is now a standard part of NEA
\citep{fath06,fath99_review}.  Subsequently, \citet{fath04_cyber} and
\citet{fath07_pyramids} found evidence that network homogenization
occurred regularly in large hypothetical ecosystem networks, and it
tended to increase with energy--matter recycling as well as network
size.  In contrast to these hypothetical models, however, there have
been comparatively few applications of this analysis to
empirically-based ecosystem models \citep[but see][]{fath99_homo,
  borrett07_lanier, gattie06, baird09_aggregation}.

The objective of the work reported here was to to determine the evidence
for the network homogenization hypothesis in empirically-based
ecosystem models.  Specifically, we tested two hypotheses.  First, we
examined the generality of the  occurrence of network homogenization
in 50 empirically-based trophic ecosystem models.  Second, we tested the
hypothesized relationship between network homogenization and network
order, connectance degree, and the magnitude of recycling.  We
concluded by considering the ecological consequences of this
type of connectivity in ecosystem networks.

\section{Materials and Methods}
\subsection{Models}
For this meta-analysis we used a database of 50 ecosystem network
models that represent 35 distinct primarily freshwater and marine
ecosystems (Table~\ref{tab:models}).  The models exhibit a range of
sizes ($4 \le n \le 125$), connectance (number of direct links divided
by the total possible number of links; $0.03 \le (C=L/n^2) \le 0.40$),
and recycling ($0 \le FCI \le 0.51$).  At the core of these
trophically-based networks is a food web; however, any
ecological processes that transfers energy--matter are considered
including natural mortality, excretion, and respiration.  Furthermore,
detritus and several other non-living pools of carbon (e.g., dissolved
organic matter) are usually key components of the models.
\input{table1}

These ecosystem models are empirically derived and generally meet the
network construction criteria of \citet{fath07_netconstruction}.  We
designate them as empirically based because empirical estimates were
made to quantify some number of the energy--matter fluxes; however,
they are heterogeneous with respect to their degree of empirical
quantification and resolution.  Sometimes to complete the model estimates were made
from similar systems or generalities, rather than the specific system
observed \citep[e.g.][]{brylinsky72}.  Despite this variability, we
claim that these models are empirically based by contrast to
non-empirically based ecosystem models such as the hypothetical models
of \citet{webster75} or the cyber-models built from general ecosystem
assembly rules like those of \citet{fath04_cyber}.

\subsection{Network Environ Analysis and Homogenization}
Our study uses Network Environ Analysis (NEA), which is well described
in the literature \citep[e.g.,][]{patten76, fath99_review, fath06}.
We introduce aspects of the output oriented throughflow analysis most
relevant to the work presented.

NEA is applied to a network model of energy--matter storage and flux
in ecosystems.  In this model, $n$ nodes represent species, groups of
species, or abiotic components and the $L$ weighted directed edges
represent the flow of energy--matter generated by some ecological
process (e.g., photosynthesis, consumption, excretion, harvesting, and
respiration).  Let $\mathbf{F}_{n\times n}=(f_{ij})$ represent the
observed flow from ecosystem compartment $j$ to compartment $i$ (e.g.,
$j \rightarrow i$), $\vec{z}_{n\times 1}$ be a vector of node inputs
originating from outside the system, and $\vec{y}_{1\times n}$ be a
vector of flows from each node that exit the system.  This network
model is like a road map for the transportation of energy or nutrients
through the ecosystem.  To apply NEA to the models, we usually assume
they are at a static, steady-state \citep[balanced inputs and outputs,
but see][for possible ways to relax these assumptions]{finn80,
  shevtsov09_dea}.  To meet this analytical assumption, we balanced
22/50 models in our data set that were not initially at steady state
with the AVG2 algorithm \citep{allesina03}.

Given this model input, our analysis starts with three main calculations.
First, we determine the throughflow vector $\vec{T}$, which is the
total amount of energy--matter flowing into or out of each node.  This
can be calculated from the model information as follows:
\begin{eqnarray}
\vec{T}^{\textrm{in}}&=&\sum_{j=1}^nf_{ij} + z_i \textrm{, and} \\
\vec{T}^{\textrm{out}}&=&\sum_{i=1}^nf_{ij} + y_j.
\end{eqnarray}
At steady state, $\vec{T}^{\textrm{ in}} = (\vec{T}^{\textrm{
    out}})^{\mathrm{T}} = \vec{T}_{n\times 1} = (T_j)$.  From this
vector, we derive the first whole-system indicator, total system
throughflow ($TST=\sum_{i=1}^n\vec{T}$). $TST$ indicates the magnitude
of flow activity in the system and is similar in concept to the gross
domestic product from economics.

The second calculation determines the \emph{direct flow
  intensities}, $\mathbf{G}_{n\times n}=(g_{ij})$ from node $j$ to
$i$.  These are
\begin{equation}\mathbf{G}=(g_{ij})=f_{ij}/T_j.\end{equation}
Notice that the elements of $\mathbf{G}$ are unitless and that the
column sums must lie between zero and unity because ecosystems
are open thermodynamic systems \citep{jorgensen99_open}.

The final step is to determine the \emph{integral flow
  intensities}, $\mathbf{N}_{n\times n}=(n_{ij})$. These $n_{ij}$ represent the
intensity of boundary flow that passes from $j$ to $i$ over all
pathways of all lengths.  These values integrate the boundary, direct,
and indirect flows, and are determined as
\begin{equation} \label{eq:G}
\mathbf{N} = \sum_{m=0}^\infty \mathbf{G}^m = \underbrace{\mathbf{I}}_{\textrm{Boundary}} + \underbrace{\mathbf{G}^1}_{\textrm{Direct}} + \underbrace{\mathbf{G}^2 + \ldots + \mathbf{G}^m + \ldots}_{\textrm{Indirect}}.
\end{equation}
In equation (\ref{eq:G}), $\mathbf{I}=(i_{ij})=\mathbf{G}^0$ is the
matrix multiplicative identity, and the elements of $\mathbf{G}^m$ are
the flow intensities from $j$ to $i$ over all pathways of length $m$.
For example, the pathway $j \rightarrow k \rightarrow i$ would have a
length $m=2$.  We can find the exact values of $\mathbf{N}$
because the power series converges so that
\begin{equation} \mathbf{N}= (\mathbf{I}-\mathbf{G})^{-1}. \end{equation}

Multiple manipulations of the $\mathbf{G}$ and $\mathbf{N}$ matrices
have been made in NEA \citep{fath06,schramski06}, but here we are most
concerned with network homogenization.  \citet{fath99_homo}
suggested we could use the ratio of the coefficients of variation
in $\mathbf{G}$ and $\mathbf{N}$ to quantify the homogenization of
resources hypothesized to occur over the longer pathways captured in
the integral matrix.  Network homogenization is
defined as
\begin{equation} 
  HMG \equiv \frac{CV(\mathbf{G})}{CV(\mathbf{N})}.  \label{eq:hmg}
\end{equation} 
$CV(\mathbf{G})$ and $CV(\mathbf{N})$ represent the standard deviation
of the elements in the matrices divided by the mean of the matrix
elements.  \citet{fath99_homo} showed algebraically that
$CV(\mathbf{N})$ scales between 0 and $n$, but $CV(\mathbf{G})$
should be less than $n$.  The $HMG$ ratio has the nice property of
indicating how much more evenly the resources are spread across the
flows when indirect pathways are considered relative to the initial
distribution in the direct flow intensity matrix.  If $HMG$ is greater
than unity, we conclude the network has homogenized the resource
distribution.  We also consider $CV(\mathbf{N})$ to compare the raw
differences in resource homogenization between the network models.

\subsection{Uncertainty Analysis}
To determine the robustness of our results, we tested the uncertainty
of the NEA homogenization parameter in response to potential error in
the data used to construct the models.  We used a Monte Carlo type
perturbation procedure for this analysis.  For each of the 50 models
in our database, we constructed a set of perturbed models by randomly
(uniform distribution) altering the values in the initial $\mathbf{F}$
matrix by $\pm 5\%$.  Following these changes, the values of the
output vector $\vec{y}$ were modified as necessary to re-balance the
models.  If a negative $\vec{y}$ value was required, the candidate
perturbed model was considered a failure because the data must be
non-negative for these ecosystem models.  Candidate models were
generated until 10,000 successful perturbed models were created.  We
verified that the variation in the successful 10,000 models was $\pm
5\%$, and then calculated the NEA homogenization parameter for these
models. To determine the impact of the $\pm 5\%$ error in the flow
values, we characterized the resultant distribution of the $HMG$
indicator. We chose not to alter the input vector $\vec{z}$ and used a
flux perturbation technique that would minimize the change to the
original model weighted degree distributions.

\section{Results}

\begin{figure*}[t]
 \center
 \includegraphics[scale=0.85]{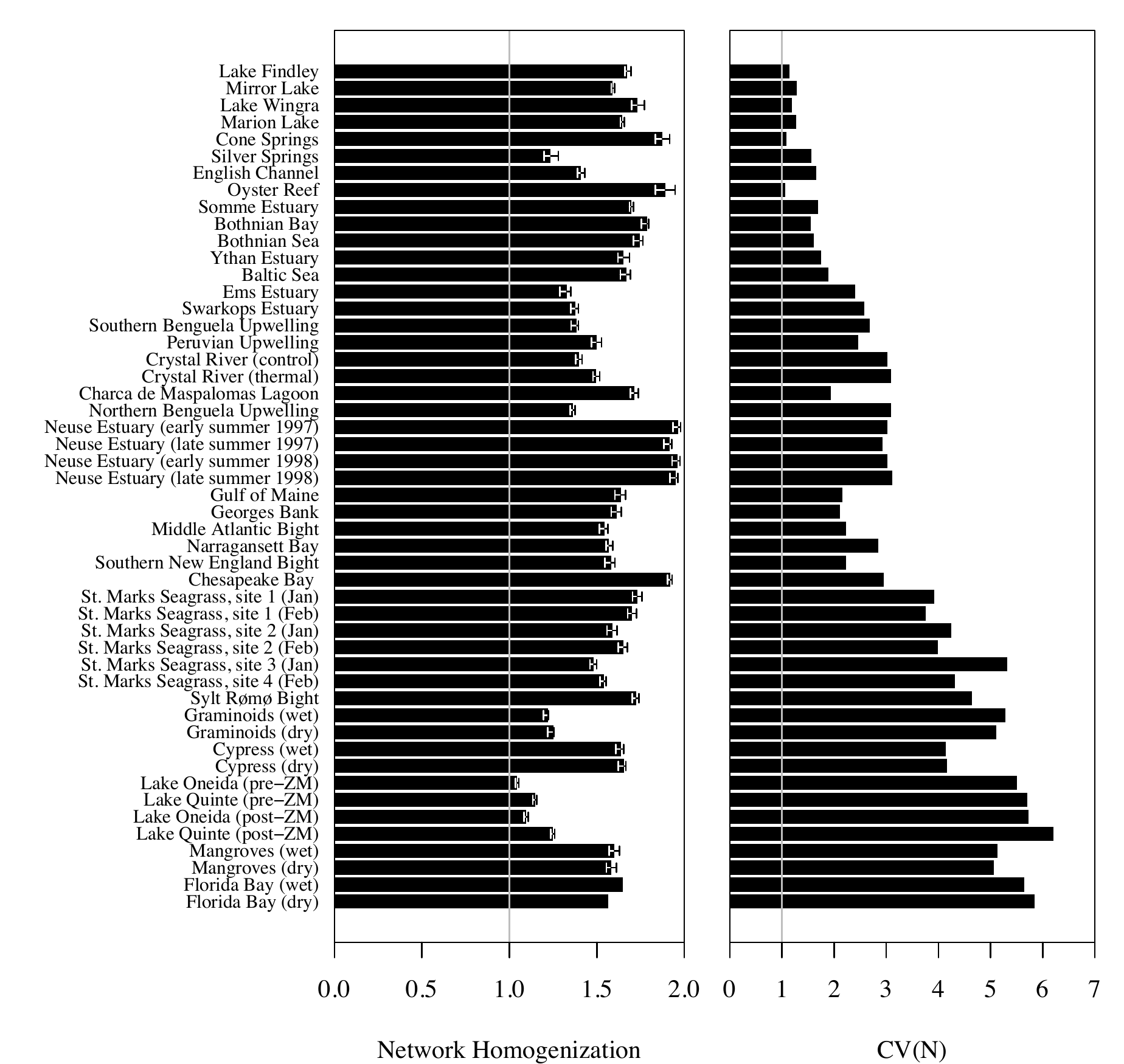}
 \caption{The degree of resource homogenization in 50
   trophically-based ecosystem models.  Panel (a) shows the network
   homogenization index $HMG$, while (b) reports the coefficient of
   variation for the elements of the integral flow intensity matrix
   $\mathbf{N}$, which is the denominator of $HMG$. Error bars in (a)
   show the range of $HMG$ values for 10,000 perturbations ($\pm 5\%$)
   of each empirically based ecosystem model.} \label{fig:hmg}
\end{figure*}

Our results provide strong evidence for the network homogenization
hypothesis.  The homogenization ratio was larger than unity in all
models analyzed (Figure~\ref{fig:hmg}a).  The minimum, median, and
maximum values are 1.04, 1.61, and 1.97, respectively.  This indicates
that the elements of the integral flow intensity matrix $\mathbf{N}$
are relatively more similar than the elements of the direct flow
intensity matrix $\mathbf{G}$ in all 50 models. Indirect flows
distribute the resource flows throughout the system.

Figure~\ref{fig:hmg}b reveals that although resource homogenization is
happening, a large amount of variability in integral flow intensities
in the models remains. The coefficient of variation in the integral
flow matrices ranged from 1.07 in the Oyster Reef model to 6.21 in the
Lake Quinte model after the zebra mussel (\textit{Dreissena
  polymorpha}) invasion.

\begin{figure*}[t] 
    \center
  \includegraphics[scale=1]{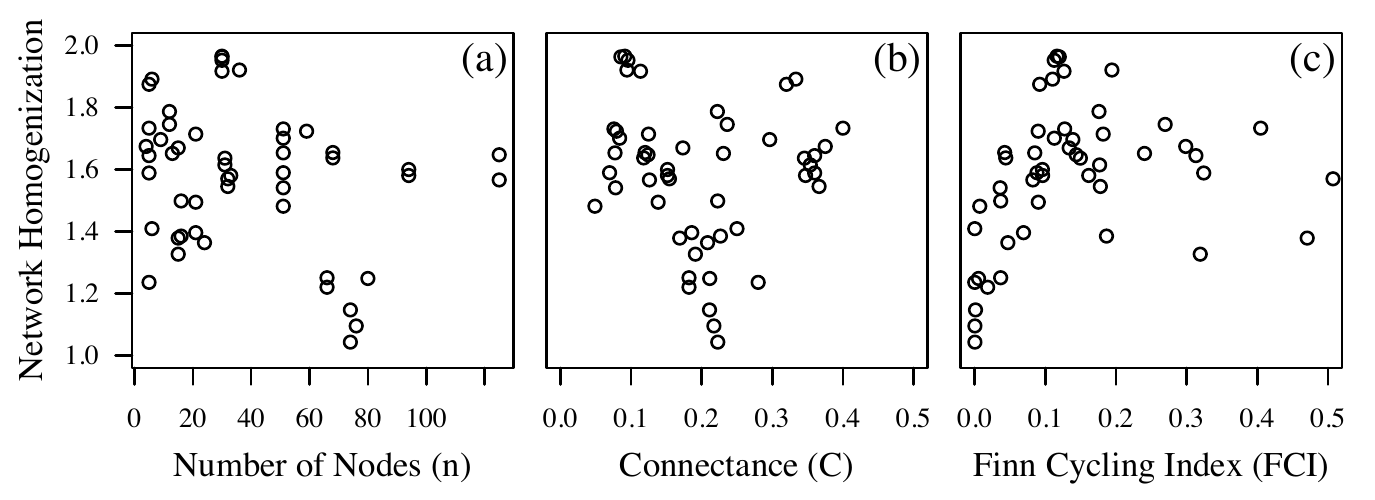}
  \caption{The relationship between output oriented network
    homogenization and (a) the number of model nodes, $n$, (b) network
    connectance, $C$, and (c) the Finn Cycling Index, $FCI$.}\label{fig:hmgV}
\end{figure*}

Figure~\ref{fig:hmgV} shows no clear relationship between the degree
of network homogenization and the model number of nodes or degree of
connectance.  Simple linear regression for $HMG$ versus model size $n$
($F = 3.65$, $r^2=0.07$, $p = 0.062$) and model connectance $C$,
($F = 0.74$, $r^2=0.02$, $p = 0.39$) are not statistically
significant. Homogenization does seem to increase with the Finn
cycling index $FCI$; however, the shape of the relationship is not
well resolved.  The relationship does not appear to be linear as a
linear regression model violates the assumption of residual
homoscedasticity.  $FCI$ and $HMG$ are significantly correlated
(Spearman's $S=11,355$, $\rho=0.45$, $p<0.001$), but this
significance disappears when we exclude the four models with an $FCI$ less than
0.01 so it may be driven by outlying data points.

\begin{figure}[t]
  \center
  \includegraphics[scale=0.8]{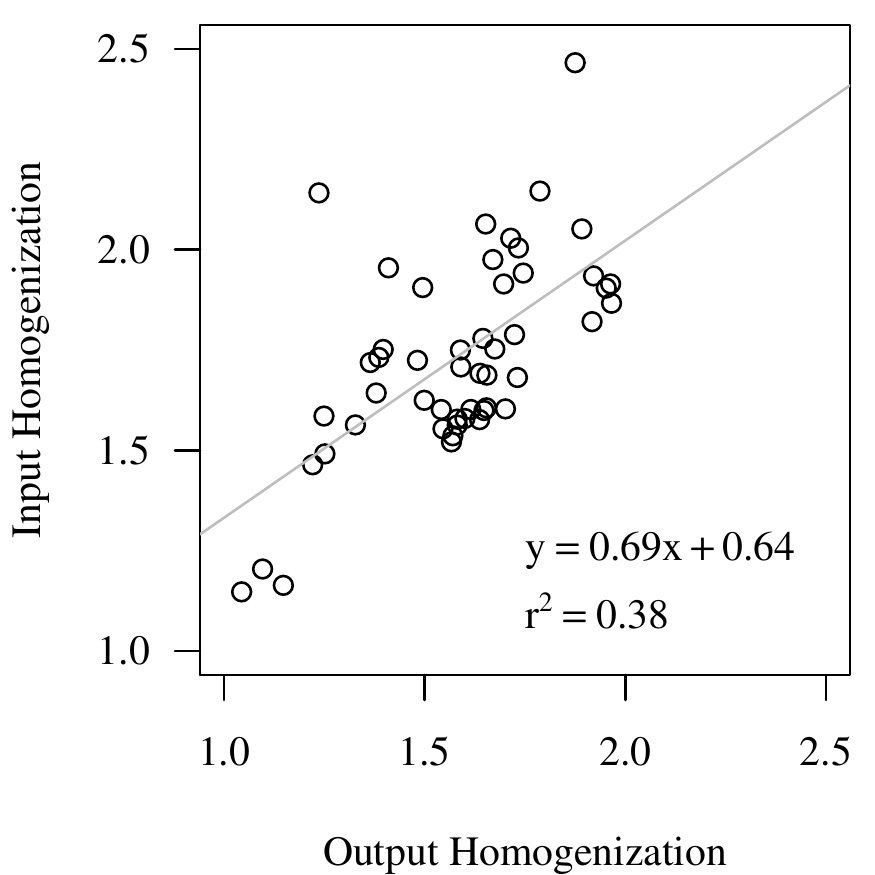}
  \caption{The relationship between input and output oriented
    calculations of network homogenization in 50 trophically-based
    ecosystem networks.  The linear regression is significant with
    $p <0.001$.}   \label{fig:hmg-inout}
\end{figure}

Many recent NEA studies have presented either the input- or
output-oriented NEA analyses in an effort to simplify the papers.  The
authors usually claim the results will be qualitatively similar
using either orientation.  Figure~\ref{fig:hmg-inout} shows this
assumption to be generally true for the network homogenization
parameter. We found that a statistically significant linear regression
fits the data (input = 0.69*output + 0.64, $p < 0.001$).  However, it
does not explain much of the variation as the adjusted $r^2$ was only
$0.38$.  Despite this, a Spearman Rank correlation test shows that the
values are statistically well correlated ($S=8,054$, $\rho = 0.61$, $p<0.001$).

Our uncertainty analysis suggests the NEA homogenization parameter is
robust to the $\pm 5\%$ variation added to the internal system fluxes
$\mathbf{F}$.  The error bars in Figure~\ref{fig:hmg} show the
distance between the minimum and maximum
of the 10,000 $HMG$ values of the perturbed models.  These ranges
tended to be small, with a median and maximum value of 0.04 and 0.11.
We were unable to generate balanced perturbed models for the Florida
Bay models because the output vector $\vec{y}$ contained 11 initial
zero values, which consistently led to negative values in the modified
$\vec{y}$ and failed perturbed models.

To further estimate the impact of the data uncertainty, we divided the
inter-quartile range (distance between first and third quartiles) by
the actual homogenization value and multiplied by 100 to get the
percent uncertainty.  The minimum, median and maximum uncertainties
for the model population was 0.24\%, 0.58\%, and 1.5\% respectively.
These values are less than the initial $\pm 5\%$ error introduced into
the $\mathbf{F}$ matrix, revealing that the error in the
homogenization parameter is generally less than the initial error.
Despite the model perturbations, the qualitative
interpretation of the homogenization parameter does not
change.  

\section{Discussion}
The core contribution of this work to systems ecology and ecological
network analysis is that we provide evidence for the generality of the
network homogenization hypothesis in empirically derived ecosystem
models. This hypothesis states that fluxes of resources are more
evenly distributed in the network when integral flow intensities
(boundary, direct, and indirect) are considered instead of just direct
flow intensities.  This was universally true in the 50
empirically derived ecosystem models we examined.  Further, we found
these results to be robust to small perturbations to the original model
flux estimates.  This suggests that small uncertainties in the
original model data would not change the outcome of our analyses.  In
this section, we discuss the context and ecological significance of
these results and conclude by outlining several possible next steps for
this research line.

\subsection{Values of Network Homogenization}
\citet{baird91} warn against comparing network indicators between
models with different structures.  This is problematic because
different model aggregation schemes as well as other structural
features can influence ecological network indicators \citep{cale79,
  abarca02, allesina05_aggregation, baird09_aggregation}.  Our goal
here is not to compare the individual values of $HMG$, but to provide
a broader context to understand the values we observed.

The $HMG$ values for the empirically derived ecosystem
models (1.04--1.97) tend to be smaller than those found in previous
studies.  \citet{fath04_cyber} found  $2< HMG\le 2.8$ in
hypothetical models based on community assembly rules. This difference
in magnitude may be because the community assembly models tended to be
larger than the empirically-based models used in this study, though
our results do not support the hypothesized relationship between $HMG$
and model size.  As suggested by \citet{fath07_pyramids}, this could
also be an artifact of the selected pattern of connections in the
assembled models.  When these authors built community assembly type
models that better matched observed food webs the range of $HMG$
values were more similar to those we present.  This is further
evidence for the importance of carefully considering the patterns of
how organisms and their environmental components are connected.  The
values previously reported for empirically derived models like the
Neuse River Estuary, 2.07 \citep{gattie06}, and Lake Sidney Lanier,
3.10 $\pm$0.31 \citep{borrett07_lanier} are also larger than the
values we found.  While there are multiple factors that may cause this
including the aggregation decisions, these two models are
biogeochemically-based networks rather than the trophically-based
models we analyzed here.  This difference may be another source of
variation \citep{baird08_sylt, borrett10_idd}.

It is surprising that our results do not provide evidence to support
the hypothesized positive relationship between $HMG$ and model size
$n$ and only weak evidence that it increases with cycling.  We suspect
this contradiction occurs because our model sample is insufficient to
show the expected pattern.  The range of model size remains
limited (Table~\ref{tab:models}).  However, if this result holds to
future scrutiny it might suggest an important characteristic of the
more empirically derived models that is not present in
\citeapos{fath04_cyber} community assembly models.

The coefficient of variation of the integral flow matrix, the
denominator in the resource homogenization ratio
(equation~\ref{eq:hmg}), shows that despite a tendency for
resource homogenization, there remains quite a bit of variability in
the integral flow intensity.  This variability increases with network
size.  This pattern is likely driven by the fact that the maximum
value of $CV(\mathbf{N})$ increases with model size
\citep[see][]{fath04_cyber}.  We don't observe the same trend with
$HMG$ because $CV(\mathbf{G})$ tends to increase with model size at
nearly the same rate observed for $CV(\mathbf{N})$.  The best fit
linear regression of $CV(\mathbf{G})$ and $n$ is
$CV(\mathbf{G})=0.07*n+1.9$ ($p<0.001, r^2=0.87$) and the linear
regression of $CV(\mathbf{N})$ and $n$ is $CV(\mathbf{N})=0.05*n+0.81$
($p<0.001, r^2=0.84$).  These trends tend to cancel in the ratio
measure.

\subsection{Ecological Significance}
Network homogenization is ecologically significant because it concerns
the distribution of resources in the system.  As stated in the
introduction, resource homogenization implies that regardless of where
the resources (e.g., carbon, nitrogen) enter the ecosystem they are
more evenly distributed in network models than we might expect from
the pattern of direct interactions.  Presence of network
homogenization suggests that the indirect flows are distributing the
model currency (energy--matter) more evenly over the links between
organisms and their environs.  Thus, the energy--matter may be more or
less available to some species than it at first appears.  This
resource homogenization may lead to a more distributed control in the
ecosystems \citep{schramski06, schramski07}, and transform the
effective relationships between species such as apparent predators
effectively operating as a net mutualist \citep{patten91,ulanowicz90}.

Resource homogenization is a consequence of the indirect
interactions in these networks.  
Indirect effects are hypothesized to generally be the dominant
component of ecological interactions \citep{higashi89,patten91,
  borrett06_neuse,borrett10_idd}, but this may not hold universally.
However, network homogenization suggests that even when
indirect effects are not dominant, they may still have important
consequences for the ecological and evolutionary interactions in the
system.  Further, we wonder if there might be a significant difference
in $HMG$ between ecosystems classified as healthy and those that are
stressed or impacted in some way.  We hypothesize that there might be
an optimal range of $HMG$ in healthy ecosystems.  Too little resource
homogenization and the system vigor or activity is under optimized,
but too much homogenization and the system becomes crystallized or
brittle in the sense of \citet{rapport98} and \citet{mageau98}.  This
possibility requires further investigation.

\subsection{Limitations and Future Work}


Further work in this research line are key to confirming the theoretical
results.  These steps can be divided into theoretical and empirical
components.

Theoretical developments are required to address the
limitations of the work we present.  The first issue concerns our
sample of ecological models.  While we have 50 models that represent
35 distinct systems, this remains a relatively small sample size that
needs to be extended as new models become available.  Furthermore,
many of the models were created by a small set of authors.  Thus, our
results could be influenced by their conscious or unconscious modeling
biases, including how they choose to aggregate species
\citep[see][]{baird09_aggregation}.  Given the universality of the
results, we do not expect this to undermine our conclusions, but it is
a potential bias that can be addressed in the future.  Another
limitation of the models is that they are all trophically-based models
of mostly aquatic ecosystems.  While we expect the results will hold
in more biogeochemically-based ecosystem models as well as more
terrestrial ecosystems, this is a testable hypothesis.

Empirical validation of Ecological Network Analysis (ENA) like the NEA
results reported here generally remains a challenge for the acceptance
and application of the theory \citep{dame06, dame08_validation}.  Part
of the challenge is that many of the ENA predictions are simply not
directly empirically testable.  However, the alignment of ENA results
like the dominance of indirect effects \citep{higashi89,
  borrett07_lanier, borrett10_idd} and empirical work
showing the importance of indirect effects \citep[e.g.,][]{menge95,
  wootton91, wootton93, menendez07, mccormick09} and initial empirical
validation attempts such as that by \citet{dame08_validation} lend
credibility to the work.  We speculate, however, that the network
homogenization hypothesis may be empirically testable using either
radio isotopes in experiments like \citet{patten67} or stable isotopes
in a manner similar to \citet{dame08_validation}.  Further, we wonder
what correspondence might exist between the network homogenization
hypothesis and stable isotope mixing models.

\subsection{Summary}
Our work provides evidence for the generality of the
systems ecology \emph{network homogenization} hypothesis in
empirically derived ecosystem models.  Homogenization was
universal in the models and the results are robust to potential
data uncertainty.  Together with the initial work of
\citet{fath99_homo} that introduced the quantification and application
of the homogenization metric and the work of \citet{fath04_cyber} and
\citet{fath07_pyramids} that investigated the phenomenon in large
ecosystem models built from community assembly rules, this is strong
evidence for the hypothesized tendency for network organization to
homogenize the distribution of resources in ecosystems.  
This is an expression of the functional consequences of connectivity
patterns in ecosystem ecology.

\section{Acknowledgments}
This research and manuscript benefited from critiques by M. Freeze,
A.\ Stapleton, and S.\ L. Fann.  We would also like to thank D. Baird, J. Link, A.\
L.\ J. Miehls and R.\ E. Ulanowicz for sharing many of their network
models with us.  Last, we gratefully acknowledge financial support
from UNCW (SRB) and the James F.\ Merritt fellowship from the UNCW
Center for the Marine Science (AKS).



\end{spacing}







\end{document}

%% file: table1.tex
\begin{table*}[T]
\caption{Fifty empirically derived trophically-based ecosystem models.} \label{tab:models}
\begin{footnotesize}
\begin{center}
\begin{tabular}{l l c c r c r}
\hline
Model & units & $n^\dagger$ & $C^\dagger$ & $TST^\dagger$ & $FCI^\dagger$ & Source \\
\hline \\[-1.5ex]
Lake Findley & gC m$^{-2}$ yr$^{-1}$ & 4 & 0.38 & 51 & 0.30 & \citet{richey78} \\
Mirror Lake & gC m$^{-2}$ yr$^{-1}$ & 5 & 0.36 & 218 & 0.32 &  \citet{richey78} \\
Lake Wingra & gC m$^{-2}$ yr$^{-1}$ & 5 & 0.40 & 1,517 & 0.40 & \citet{richey78} \\
Marion Lake & gC m$^{-2}$ yr$^{-1}$ & 5 & 0.36 & 243 & 0.31 & \citet{richey78} \\
Cone Springs & kcal m$^{-2}$ yr$^{-1}$ & 5 & 0.32 & 30,626 & 0.09 & \citet{tilly68} \\
Silver Springs & kcal m$^{-2}$ yr$^{-1}$ & 5 & 0.28 & 29,175 & 0.00 & \citet{odum57} \\
English Channel & kcal m$^{-2}$ yr$^{-1}$ & 6 & 0.25 & 2,280 & 0.00 & \citet{brylinsky72} \\
Oyster Reef & Kcal m$^{-2}$ yr$^{-1}$ & 6 & 0.33 & 84 & 0.11 & \citet{dame81} \\
Somme Estuary & mgC m$^{-2}$ d$^{-1}$ & 9 & 0.30 & 2,035 & 0.14 & \citet{rybarczyk03} \\
Bothnian Bay & gC m$^{-2}$ yr$^{-1}$ & 12 & 0.22 & 130 & 0.18 &  \citet{sandberg00} \\
Bothnian Sea & gC m$^{-2}$ yr$^{-1}$ & 12 & 0.24 & 458 & 0.27 &  \citet{sandberg00} \\
Ythan Estuary & gC m$^{-2}$ yr$^{-1}$ & 13 & 0.23 & 4,181 & 0.24 & \citet{baird81} \\
Baltic Sea & mgC m$^{-2}$ d$^{-1}$ & 15 & 0.17 & 1,974 & 0.13 &  \citet{baird91} \\
Ems Estuary & mgC m$^{-2}$ d$^{-1}$ & 15 & 0.19 & 1,019 & 0.32 & \citet{baird91} \\
Swarkops Estuary & mgC m$^{-2}$ d$^{-1}$ & 15 & 0.17 & 13,996 & 0.47 &  \citet{baird91} \\
Southern Benguela Upwelling & mgC m$^{-2}$ d$^{-1}$ & 16 & 0.23 & 1,774 & 0.19 &\citet{baird91} \\
Peruvian Upwelling & mgC m$^{-2}$ d$^{-1}$ & 16 & 0.22 & 33,496 & 0.04 & \citet{baird91} \\
Crystal River (control) & mgC m$^{-2}$ d$^{-1}$ & 21 & 0.19 & 15,063 & 0.07 & \citet{ulanowicz86, ulanowicz95} \\
Crystal River (thermal) & mgC m$^{-2}$ d$^{-1}$ & 21 & 0.14 & 12,032 & 0.09 & \citet{ulanowicz86,ulanowicz95} \\
Charca de Maspalomas Lagoon & mgC m$^{-2}$ d$^{-1}$ & 21 & 0.13 & 6,010,331 & 0.18 & \citet{alumnia99} \\
Northern Benguela Upwelling & mgC m$^{-2}$ d$^{-1}$ & 24 & 0.21 &6,608 & 0.05 & \citet{heymans00} \\
Neuse Estuary (early summer 1997) & mgC m$^{-2}$ d$^{-1}$ & 30 & 0.09 & 13,826 & 0.12 & \citet{baird04} \\
Neuse Estuary (late summer 1997) & mgC m$^{-2}$ d$^{-1}$ & 30 & 0.11 & 13,038 & 0.13 & \citet{baird04} \\
Neuse Estuary (early summer 1998) & mgC m$^{-2}$ d$^{-1}$ & 30 & 0.09 & 14,025 & 0.12 & \citet{baird04} \\
Neuse Estuary (late summer 1998) & mgC m$^{-2}$ d$^{-1}$ & 30 & 0.10 & 15,031 & 0.11 & \citet{baird04} \\
Gulf of Maine & g ww m$^{-2}$ yr$^{-1}$ & 31 & 0.35 & 18,382 & 0.15 &  \citet{link08} \\
Georges Bank & g ww m$^{-2}$ yr$^{-1}$ & 31 & 0.35 & 16,890 & 0.18 & \citet{link08} \\
Middle Atlantic Bight & g ww m$^{-2}$ yr$^{-1}$ & 32 & 0.37 & 17,917 & 0.18 & \citet{link08} \\
Narragansett Bay & mgC m$^{-2}$ yr$^{-1}$ & 32 & 0.15 & 3,917,246 & 0.51  & \citet{monaco97} \\
Southern New England Bight & g ww m$^{-2}$ yr$^{-1}$ & 33 & 0.03 & 17,597 & 0.16 & \citet{link08} \\
Chesapeake Bay  & mgC m$^{-2}$ yr$^{-1}$ & 36 & 0.09 & 3,227,453 & 0.19 & \citet{baird89} \\
St. Marks Seagrass, site 1 (Jan) & mgC m$^{-2}$ d$^{-1}$ & 51 & 0.08 & 1,316 & 0.13 & \citet{baird98} \\
St. Marks Seagrass, site 1 (Feb) & mgC m$^{-2}$ d$^{-1}$ & 51 & 0.08 & 1,591 & 0.11 & \citet{baird98} \\
St. Marks Seagrass, site 2 (Jan) & mgC m$^{-2}$ d$^{-1}$ & 51 & 0.07 & 1,383 & 0.09 & \citet{baird98} \\
St. Marks Seagrass, site 2 (Feb) & mgC m$^{-2}$ d$^{-1}$ & 51 & 0.08 & 1,921 & 0.08 & \citet{baird98}\\
St. Marks Seagrass, site 3 (Jan) & mgC m$^{-2}$ d$^{-1}$ & 51 & 0.05 & 12,651 & 0.01 &\citet{baird98}\\
St. Marks Seagrass, site 4 (Feb) & mgC m$^{-2}$ d$^{-1}$ & 51 & 0.08 & 2,865 & 0.04 & \citet{baird98}\\
Sylt R{\o}m{\o} Bight & mgC m$^{-2}$ d$^{-1}$ & 59 & 0.08 & 1,353,406 & 0.09 & \citet{baird04_sylt} \\
Graminoids (wet) & gC m$^{-2}$ yr$^{-1}$ & 66 & 0.18 & 13,677 & 0.02 & \citet{ulanowicz00_graminoids} \\
Graminoids (dry) & gC m$^{-2}$ yr$^{-1}$ & 66 & 0.18 & 7,520 & 0.04 &  \citet{ulanowicz00_graminoids} \\
Cypress (wet) & gC m$^{-2}$ yr$^{-1}$ & 68 & 0.12 & 2,572 & 0.04 & \citet{ulanowicz97_cypress} \\
Cypress (dry) & gC m$^{-2}$ yr$^{-1}$ & 68 & 0.12 & 1,918 & 0.04 & \citet{ulanowicz97_cypress}\\
Lake Oneida (pre-ZM) & gC m$^{-2}$ yr$^{-1}$ & 74 & 0.22 & 1,638 & $<0.01$ & \citet{miehls09_oneida} \\
Lake Quinte (pre-ZM) & gC m$^{-2}$ yr$^{-1}$ & 74 & 0.21 & 1,467 & $<0.01$ &  \citet{miehls09_quinte} \\
Lake Oneida (post-ZM) & gC m$^{-2}$ yr$^{-1}$ & 76 & 0.22 & 1,365 & $<0.01$ & \citet{miehls09_oneida} \\
Lake Quinte (post-ZM) & gC m$^{-2}$ yr$^{-1}$ & 80 & 0.21 & 1,925 & $0.01$ &  \citet{miehls09_quinte} \\
Mangroves (wet) & gC m$^{-2}$ yr$^{-1}$ & 94 & 0.15 & 3,272 & 0.10 & \citet{ulanowicz99_mangrove} \\
Mangroves (dry) & gC m$^{-2}$ yr$^{-1}$ & 94 & 0.15 & 3,266 & 0.10 & \citet{ulanowicz99_mangrove} \\
Florida Bay (wet) & mgC m$^{-2}$ yr$^{-1}$ & 125 & 0.12 & 2,721 & 0.14 & \citet{ulanowicz98_fb} \\
Florida Bay (dry) & mgC m$^{-2}$ yr$^{-1}$ & 125 & 0.13 & 1,779 & 0.08& \citet{ulanowicz98_fb} \\[0.5ex] \hline\\[-1.5ex]
\end{tabular}
\end{center}
\end{footnotesize}
\begin{scriptsize}

$^\dagger$ $n$ is the number of nodes in the network model, $C=L/n^2$
is the model connectance when $L$ is the number of direct links or
energy--matter transfers, $TST=\sum\sum{f_{ij}}+\sum{z_i}$ is the total
system throughflow, and $FCI$ is the Finn Cycling Index.  
\end{scriptsize}
\end{table*}